\documentclass[twocolumn,aps,showpacs,prb,amsmath,amssymb,floatfix,superscriptaddress]{revtex4-2}

\usepackage{amssymb,color}

\usepackage{bm} 
\usepackage{color}
\usepackage{graphicx}
\usepackage{physics}
\usepackage{amsthm}
\usepackage{amsmath}
\usepackage{amssymb}
\usepackage{enumerate}
\usepackage{placeins}
\usepackage{booktabs}
\usepackage{dsfont}
\usepackage{yfonts}
\usepackage{hyperref} 
\usepackage{epsfig} 
\usepackage{float}

\DeclareSymbolFont{bbold}{U}{bbold}{m}{n}
\DeclareSymbolFontAlphabet{\mathbbold}{bbold}

\newcommand{\e}{{\rm e}}  
	
\newcommand{\up}{\uparrow}	
\newcommand{\down}{\downarrow}

\newcommand{\td}{\mathrm{d}}
\newcommand{\Eso}{E_{\rm so}}
\newcommand{\lso}{l_{\rm so}}

\begin{document}

\title{Topological gap protocol based machine learning optimization of Majorana hybrid wires}

\author{Matthias Thamm}
\affiliation{Institut f\"{u}r Theoretische Physik, Universit\"{a}t Leipzig,  Br\"{u}derstrasse 16, 04103 Leipzig, Germany} 
\author{Bernd Rosenow}
\affiliation{Institut f\"{u}r Theoretische Physik, Universit\"{a}t Leipzig,  Br\"{u}derstrasse 16, 04103 Leipzig, Germany}

\date{\today}

\begin{abstract}
    Majorana zero modes in superconductor-nanowire hybrid structures are a promising candidate for topologically protected qubits with the potential to be used in scalable structures. Currently, disorder in such Majorana wires is a major challenge as it can destroy the topological phase and thus reduce  the yield in the fabrication of Majorana devices. We study machine learning optimization of a gate array in proximity to a grounded Majorana wire, which allows us to reliably compensate even strong disorder. We propose a metric for optimization that is inspired by the topological gap protocol, and which can be implemented based on measurements of the non-local  conductance through the wire.
\end{abstract}
 
\maketitle

\section{Introduction}

A promising avenue towards achieving scalable quantum computing involves the utilization of Majorana zero modes (MZMs) \cite{Ladd.2010,Kloeffel.2013,Vandersypen.2017}, which emerge as bound states within topological superconductors \cite{Alicea.2011,Clarke.2011,Hyart.2013,Sarma.2015,Aasen.2016,Karzig.2017,Lutchyn.2018,Oreg.2020,Mandal.2023}. Their occurrence is a consequence of the topological properties of the underlying phase, and they manifest as zero-energy states located within the excitation gap of the system. Due to this topological protection, MZMs exhibit robustness against external perturbations and decoherence. Furthermore, the ability to manipulate MZMs through anyonic braiding allows for the implementation of  fault-tolerant qubit operations \cite{Alicea.2011,Clarke.2011,Hyart.2013,Sarma.2015,Vijay.2016}.

MZMs can occur in hybrid systems of conventional superconductors and semiconductors with strong spin-orbit coupling \cite{Kitaev.2001,Lutchyn.2010,Oreg.2010,Sau.2010,Alicea.2011}. However, disorder in these systems turns out to be a major problem \cite{Zhang.2017,Ahn.2021,DasSarma.2021,Yu.2021,Aghaee.2022,DasSarma.2023}, as it can destroy the topological phase \cite{Takei.2013} and induce trivial Andreev bound states (ABSs) \cite{Kells.2012,Prada.2012,Rainis.2013, Cayao.2015,San.2016,Chen.2017,Liu.2017,Penaranda.2018,Avila.2019,Chiu.2019,Chen.2019,Woods.2019,Vuik.2019,awoga2019supercurrent,Dmytruk.2020,Pan.2020,Prada.2020,Valentini.2021,Zhang.2021}, which can mimic signatures of MZMs \cite{Bagrets.2012,Pikulin.2012,Liu.2012,Zhang.2017,Pan.2020,DasSarma.2021}. These ABSs complicate the verification of MZMs in experiments as they make more complex measurements and devices necessary \cite{Whiticar.2020,Aghaee.2022}. To distinguish MZMs from ABSs, signatures based on coherent transport using electron interferometers \cite{Fu.2010,Hell.2018,Whiticar.2020,Thamm.2021} are suitable, but experimentally challenging \cite{Whiticar.2020}. Another method to detect MZMs is the so-called topological gap protocol \cite{Pikulin.2021}, which can be applied to a grounded wire contacted with leads at both ends. Here, all elements of the conductance matrix between the two leads are measured to ensure that zero-bias conductance peaks occur simultaneously at both ends  and that the excitation gap closes at the boundaries of the topological phase \cite{Pikulin.2021}.

Numerous experimental studies have confirmed the predicted signatures of Majorana zero modes (MZMs) \cite{Mourik.2012,Das.2012,Deng.2012,Nichele.2017,Rokhinson.2012,Albrecht.2016}. Compelling evidence exists for the presence of MZMs in hybrid wires, demonstrated through interferometry \cite{Whiticar.2020} and the application of the topological gap protocol \cite{Aghaee.2022}. Theoretical investigations have also identified strategies for enhancing MZMs in clean Majorana wires, including the use of magnetic field textures \cite{Klinovaja.2012,Boutin.2018,Mohanta.2019,Turcotte.2020}, harmonic potential profiles \cite{Boutin.2018}, and optimized geometries for Majorana Josephson junctions \cite{Melo.2022}. However, the presence of disorder in the fabrication process significantly affects the yield of Majorana devices \cite{Aghaee.2022}, which poses a major limitation for the realization of large-scale qubit systems. Several approaches have been proposed to address this challenge. Firstly, efforts have been made to fabricate cleaner wires by improving fabrication processes \cite{Gul.2015,Sarma.2015}. Additionally, it has been demonstrated that weak coupling between superconductors and semiconductors can enhance the resilience of hybrid wires to moderate disorder \cite{Awoga.2022,Awoga.2022b}. Another strategy involves utilizing machine learning optimization techniques to create a potential profile along the Majorana wire using a gate array, compensating for the effects of disorder \cite{Thamm.2023}. However, the optimization process requires the measurement of coherent transport through an electron interferometer \cite{Fu.2010,Hell.2018,Whiticar.2020,Thamm.2021}, which may pose challenges for scalability.

\begin{figure}[t!]
    \centering
    \includegraphics[width=8.6cm]{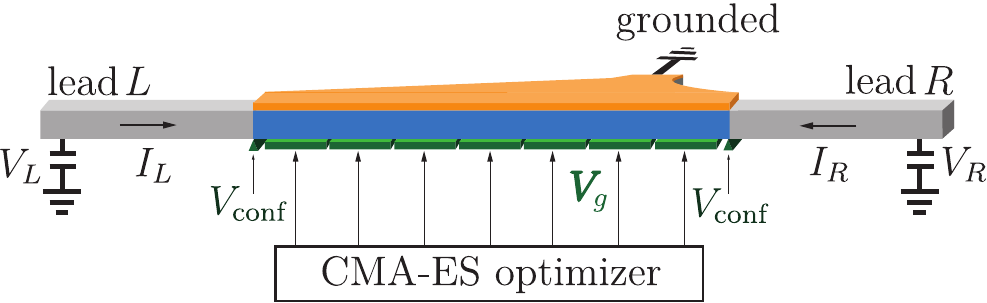}  
    \caption{\label{Fig:setup}Majorana hybrid wire consisting of a grounded superconductor (orange) and a semiconductor with strong spin orbit coupling (blue) connected to two leads $L$ and $R$ separated from the wire by a potential $V_{\rm conf}$ created by pinch-off gates. The full conductance matrix $G_{\alpha\beta}=\td I_\alpha/\td V_{\beta}$ can be measured as a function of an applied bias voltage $V_R-V_L$ and external Zeeman field $E_z$ based on which voltages of an array of gates (green) are optimized using the CMA-ES algorithm \cite{Hansen.2016} to cancel disorder effects in the hybrid wire. }
\end{figure}
In recent years, the increasing complexity of quantum devices \cite{Karzig.2017,Arute.2019,Lennon.2019,Oreg.2020,Ares.2021} has necessitated the automatic tuning of parameters \cite{Baart.2016,Botzem.2018,Kalantre.2019,Teske.2019,Mills.2019,Durrer.2020,Moon.2020,vanEsbroeck.2020,Craig.2021,Fedele.2021,Ziegler.2021,Krause.2022}. Machine learning algorithms have emerged as effective tools for this purpose \cite{Frees.2019,Kalantre.2019,Moon.2020,RuizEuler.2020,Ares.2021,Craig.2021,Ziegler.2021}. In this paper, we propose an optimization approach that employs the CMA-ES machine learning algorithm \cite{Hansen.2016} to tune the voltages on an array of gates located near a Majorana wire. Our optimization metric is based on the principle of the topological gap protocol, eliminating the need for interferometry. We demonstrate the effectiveness of the machine learning algorithm in finding gate voltages that minimize the metric, resulting in the reliable restoration of the topological phase, localized MZMs, and the topological gap, even in the presence of significant disorder. Notably, the optimization process effectively disregards trivial ABSs and can drive the wire from a trivial phase to a topological phase while mitigating the effects of disorder. Additionally, we discuss the convergence properties of the algorithm and argue that this optimization approach is experimentally feasible using currently available technology.

\section{Setup}
We consider  a grounded Majorana hybrid wire of length $L$ that is connected to two leads (labeled $L$ and $R$) at the ends  (Fig.~\ref{Fig:setup}).  An array of gates is placed in proximity to the wire such that the voltages on the individual gates can be controlled by the CMA-ES algorithm. Two additional gates that are not included in the optimization are used to separate the wire from the leads by creating a confinement potential. 

This setup allows to measure the entries of the conductance matrix $G_{\alpha\beta}=\td I_\alpha /\td V_\beta$ between the leads 
\begin{align}
    G &= \frac{e^2}{h} \left(
        \begin{matrix}
            \tilde{G}_{LL} & \tilde{G}_{LR} \\
            \tilde{G}_{RL} & \tilde{G}_{RR} 
    \end{matrix} \right) \ .
\end{align}
Denoting the voltage at lead $\alpha$ as $V_\alpha$, the entries in units of $e^2/h$ at zero temperature are given by \cite{Danon.2020}
\begin{align}
    \begin{split}
        \tilde{G}_{LL} &= N_L - R_L^e(-eV_L) + A_L^e(-eV_L)\\
        \tilde{G}_{LR} &= -T_{LR}^e(-eV_R)+A_{LR}^e(-eV_R)\\
        \tilde{G}_{RL} &= -T_{RL}^e(-eV_L)+A_{RL}^e(-eV_L) \\
        \tilde{G}_{RR} &= N_R - R_R^e(-eV_R) + A_R^e(-eV_R) \ .
    \end{split} 
\end{align} 
Here, $N_\alpha$ is the number of channels in lead $\alpha$, $R_{\alpha}^e$ is the reflection probability for an electron at lead $\alpha$, $A_{\alpha}^e$ is the probability for an electron to be reflected as a hole at lead $\alpha$, $T_{\alpha'\alpha}^e$ is the transmission probability for an electron from lead $\alpha$ to lead $\alpha'$, and $A_{\alpha'\alpha}^e$ is the probability for an electron from lead $\alpha$ to be transmitted as a hole to lead $\alpha'$ \cite{Danon.2020}.  

Below, we construct a metric for optimization based on conductance measurements. To assess whether optimization of the metric successfully restores an extended topological phase, we consider the \emph{scattering invariant} \cite{Akhmerov.2011}
\begin{align}
    \mathcal{Q} =  {\rm sgn}\; {\rm det} \left(
        \begin{matrix}
        \bm{r}_L^{e\to e} & \bm{r}_L^{e\to h}\\
        \bm{r}_L^{h\to e} & \bm{r}_L^{h\to h}
        \end{matrix}
        \right) \  .
\end{align}
This is the sign of the determinant of the reflection block obtained from the scattering matrix for lead $L$, containing the complex reflection amplitudes. If the system is in the topological phase  $\mathcal{Q}=-1$ holds, and in the trivial phase $\mathcal{Q}=+1$. However, it is important to note that this scattering invariant is not experimentally accessible and is only reported as a benchmark for assessing the performance of the optimization process for restoring the topological phase. 

In order to use MZMs as qubits in a quantum computer, it is not only necessary that the wire is in the topological phase but also that the excitation gap is large. We combine both properties in the so-called \emph{topological gap} $\mathcal{Q}\Delta_{\rm gap}$ \cite{Aghaee.2022}, where $\Delta_{\rm gap} =  \mathcal{E}_1$ is the energy of the second level, i.e.\@ the energy of the level above the MZM in the topological phase, which is approximately equal to the excitation gap. To realize qubits, a negative value for the topological gap with large magnitude is advantageous. 

For our simulations, we first study an effective one-dimensional wire and consider a more realistic two-dimensional system later. The one-dimensional Majorana wire in Nambu basis $\left(d_\up^\dagger(y),d_\down^\dagger(y),d_\down(y),	-d_\up(y)\right)$ is modeled by the BdG Hamiltonian 
\begin{align}
    \mathcal{H}_{\rm wire} = \tau_z \Bigg[ 
                &-\frac{\hbar^2\partial_y^2}{2m^*}\sigma_0 - \mu\sigma_0 
                -i\hbar\alpha_R\sigma_x\partial_y 
                \notag\\
                &+ \delta_{\rm dis}(y)\sigma_0 + V_g(y)\sigma_0+V_{\rm conf}(y) \sigma_0
                \Bigg] 
                \notag\\
                &-E_z\tau_0\sigma_z + \Delta\tau_x\sigma_0
                \ , \label{Eq:Hwire}
\end{align}
where $\hbar\alpha_R=0.2\rm\,eV\,\AA$ is the Rashba spin-orbit coupling strength and $m^*=0.02\,m_e$ is the effective mass of the electrons such that an characteristic energy scale is given by $\Eso=\alpha_R^2 m^*/2 = 0.05\,\rm meV$ and a characteristic length scale is $\lso=\hbar/(\alpha_R m^*)=0.19\,\rm \text{\textmu} m$ -- realistic values for InAs nanowires \cite{Mourik.2012,Lutchyn.2018}. Here, $\Delta=2\,\Eso$ is the proximity induced s-wave superconducting gap, $\mu$ the chemical potential, $E_z$ the Zeeman energy due to an external magnetic field $\bm{B}=B\bm{e}_z$, and $\sigma_i$ and $\tau_i$ are Pauli matrices acting in spin and particle-hole space, respectively.  Disorder in the wire is described by normally distributed random numbers $\delta_{\rm dis}(y)$ with standard deviation $\sigma_{\rm dis}$, and a finite correlation length $\lambda_{\rm dis}$ can be introduced by damping high Fourier modes (for details see Appendix A). The confinement potential $V_{\rm conf}$ separating the wire form the leads is given by a steep Gaussian peak at both lead-wire interfaces (see Appendix A). Only when considering tuning a wire from the ABSs regime to the topological phase, we choose a more shallow confinement. 

\begin{figure}[t!]
    \centering
    \includegraphics[width=8.6cm]{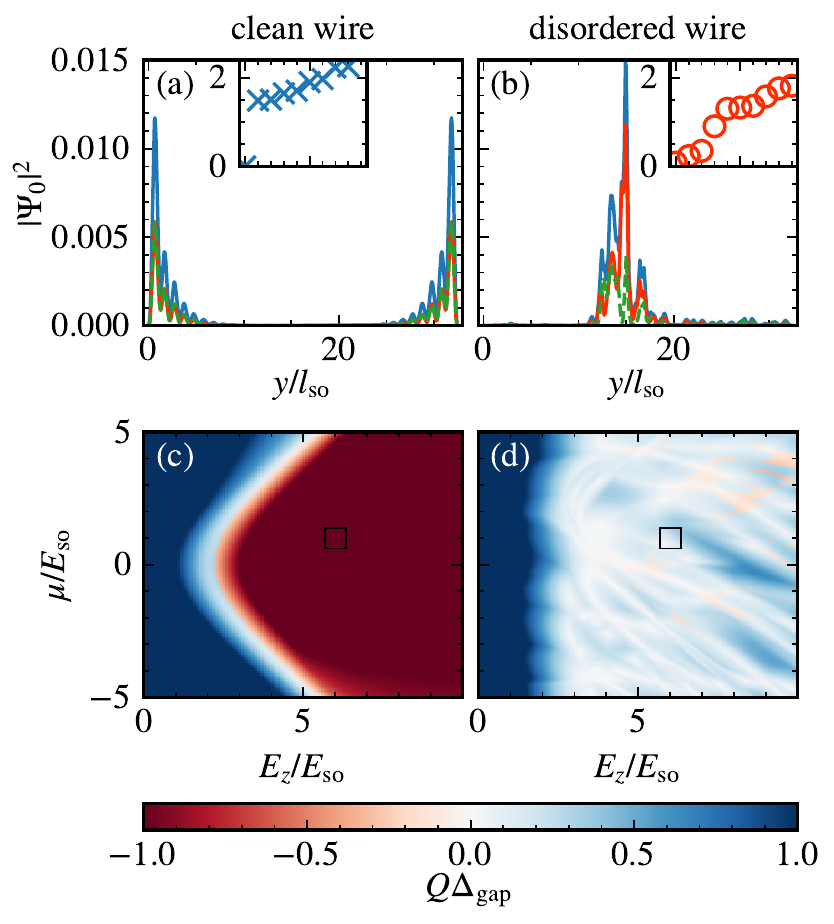}  
    \caption{\label{Fig:ref}Effects of strong disorder on a Majorana wire of length $L=32.5\,\lso$ in the topological phase for $\mu=1\,\Eso$ and $E_z=6\,\Eso$. Panels (a) and (b) depict the wave functions $|\Psi_0|^2$ (blue) of the ground state together with the electron $|u_0|^2$ (green) and hole wave function $|v_0|^2$ (orange) for a clean and disordered wire ($\sigma_{\rm dis}=25\,\Eso$, $\lambda_{\rm dis}=0$), respectively. Panels (c) and (d) show the corresponding topological gaps $\mathcal{Q}\Delta_{\rm gap}$, where $\mathcal{Q}$ is the scattering invariant and $\Delta_{\rm gap}$ is an estimator for the gap given by the energy of the second level. While the clean wire shows localized Majorana zero modes with a large topological gap ($Q=-1$, $\Delta_{\rm gap}>1\,\Eso$), both the localized zero modes and the topological phase are destroyed by strong disorder.}
\end{figure}

We use the CMA-ES algorithm (for details see Appendix B), which has found many applications for high dimensional optimization problems \cite{Hansen.2006,Lozano.2006,Loshchilov.2016,WilljuiceIruthayarajan.2010,Loshchilov.2013},  to optimize   the Fourier components $a_i, b_i$ of the $N_g$ gate voltages. The Fourier components are related to the voltage  $V_j$ on gate $j$ via 
\begin{align} 
    {V}_j = \frac{b_0}{2} 
        &+ \sum_{k=1}^{\left\lfloor \frac{N_{ g}-1}{2}\right\rfloor}
        a_k \sin\left(\frac{2\pi}{N_{\rm g}}\, k j\right) + \sum_{k=1}^{\left\lfloor \frac{N_{ g}}{2}\right\rfloor}
        b_k \cos\left(\frac{2\pi}{N_{\rm g}}\, k j\right)
          . \label{Eq:VjFourier}
\end{align} 
For reasons of better comparability, we set $b_0=0$ with the exception of the case where the optimization algorithm tunes from ABSs to MZMs. We assume that the wire is located a distance $z_{\rm sys}=0.3\,\Eso$ above the gates such that the potential created by the gate array at position $y$ in the wire is given by 
\begin{align}
    V_g(y) = \mathcal{F}^{-1}\left[\e^{-|q|z_{\rm sys}} 
        \mathcal{F}\left[\sum_{j=1}^{N_{\rm g}} {V}_j\chi_j(y)\right]
        \right] \ . \label{Eq:Vg}
\end{align}
Here, $\mathcal{F}$ and $\mathcal{F}^{-1}$ denote Fourier transform and inverse Fourier transform, respectively, and $\chi_j(y)$ is one for $y$ above gate $j$ and zero otherwise.

For computing the conductance, the eigenstates, and the energy eigenvalues of the Hamiltonian Eq.~\eqref{Eq:Hwire}, we discretize the Hamiltonian on a lattice with spacing $a=0.026\,\lso$, and use the \texttt{python} package \texttt{KWANT} \cite{Groth.2014} to extract the scattering matrix, the conductance matrix, and the Hamiltonian matrix. As a point of reference, we consider a clean wire in Fig.~\ref{Fig:ref}, for which the topological gap shows an extended topological phase as a function of Zeeman field and chemical potential for $E_z^2 > \mu^2 + \Delta^2$ (red region in panel c) with a gap closing at the boundary (white region in panel c). In the topological phase, there exists a zero energy state in the gap (inset of panel a) with a wave function localized at the wire ends (panel a). For this Majorana wave function $\Psi_0 = (\bm{u}_0,\bm{v}_0)$, the Majorana condition $|\bm{v}_0(y)|=|\bm{u}_0(y)|$ holds between electron ($\bm{u}_0=(u_\up, u_\down)$ in green) and  hole components ($\bm{v}_0$ in orange). However, when adding strong onsite disorder with standard deviation $\sigma_{\rm dis}=25\,\Eso$, the topological phase and the gap are destroyed (panel d). The eigenstate with the smallest energy is no longer localized at the wire ends and does not fulfill the Majorana condition anymore. To compensate the effects of such disorder,  in the following, we first introduce a metric and then minimize it by letting the CMA-ES algorithm find optimal gate voltages. As we will demonstrate, this optimization is capable of restoring the topological phase and the Majorana zero mode.

\section{Metric based on the topological gap protocol}\vspace{-0.4cm}
\begin{figure}[t!]
    \centering
    \includegraphics[width=8.6cm]{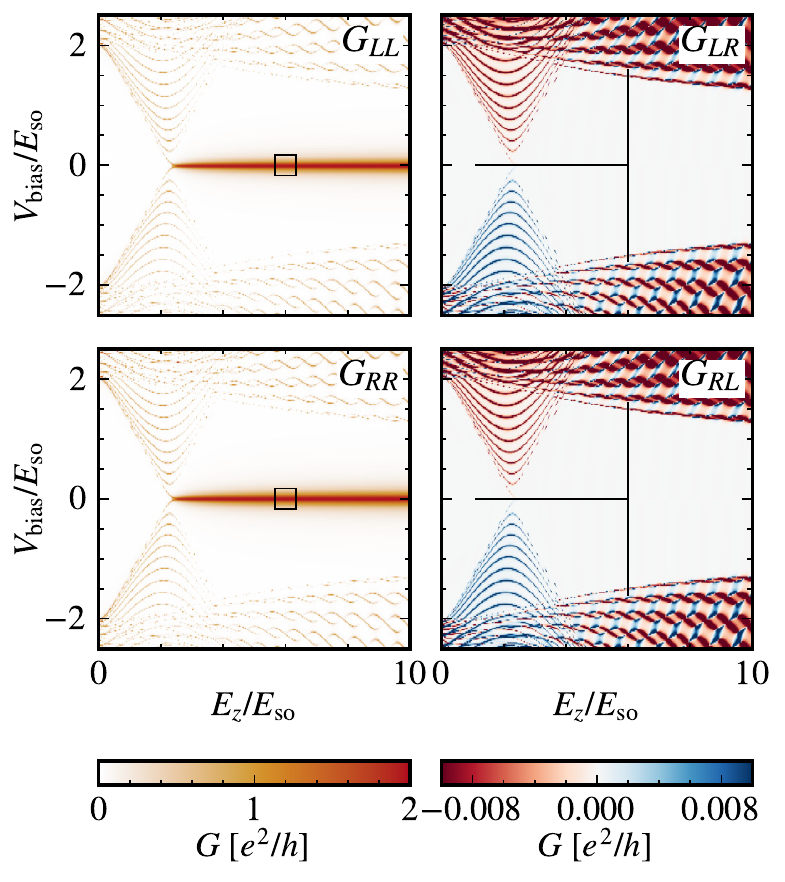}  
    \caption{\label{Fig:refmetric}Elements of the conductance matrix between lead $L$ and $R$ as a function of the Zeeman field $E_z$ and bias voltage $V_{\rm bias}$ at a chemical potential $\mu=1\,\Eso$ in a clean wire of length $L=32.5\,\lso$. For evaluating the metric, a measurement of $G_{RR}$ and $G_{LL}$ at zero bias is performed (black square), and a scan of the non-local conductances along $E_z$ for zero bias and along $V_{\rm bias}$ for a given $E_z$ are needed (black lines). Increasing the Zeeman field $E_z$, the wire enters the topological phase which shows zero bias peaks in the local conductance at both leads. At the transition, the gap closes which can be inferred from the non-local conductance.}
\end{figure}
 \begin{figure}[t!]
    \centering
    \includegraphics[width=8.6cm]{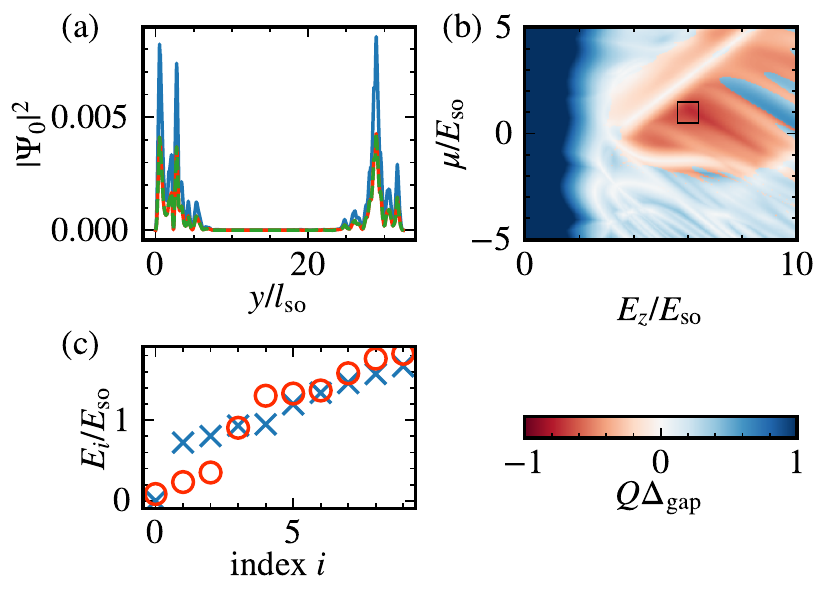}  
    \caption{\label{Fig:opt}Majorana wire with strong disorder (cf.~Fig.~\ref{Fig:ref}b) using optimized gate voltages found by minimizing the metric Eq.~\eqref{Eq:metric} with the CMA-ES algorithm. (a) Wave function $|\Psi_0|^2$ (blue) of the first level together with the electron $|\bm{u}_0|^2$ (green) and hole wave function $|\bm{v}_0|^2$ (orange), (b) topological gap $\mathcal{Q}\Delta_{\rm gap}$ , and (c) the energy levels for the optimized wire (blue) and for comparison for the disordered wire with zero voltage on all gates (red). Using optimized gate voltages  restores localized MZMs, the topological phase, and a topological gap.}
\end{figure}
\begin{figure}[t!]
    \centering
    \includegraphics[width=8.6cm]{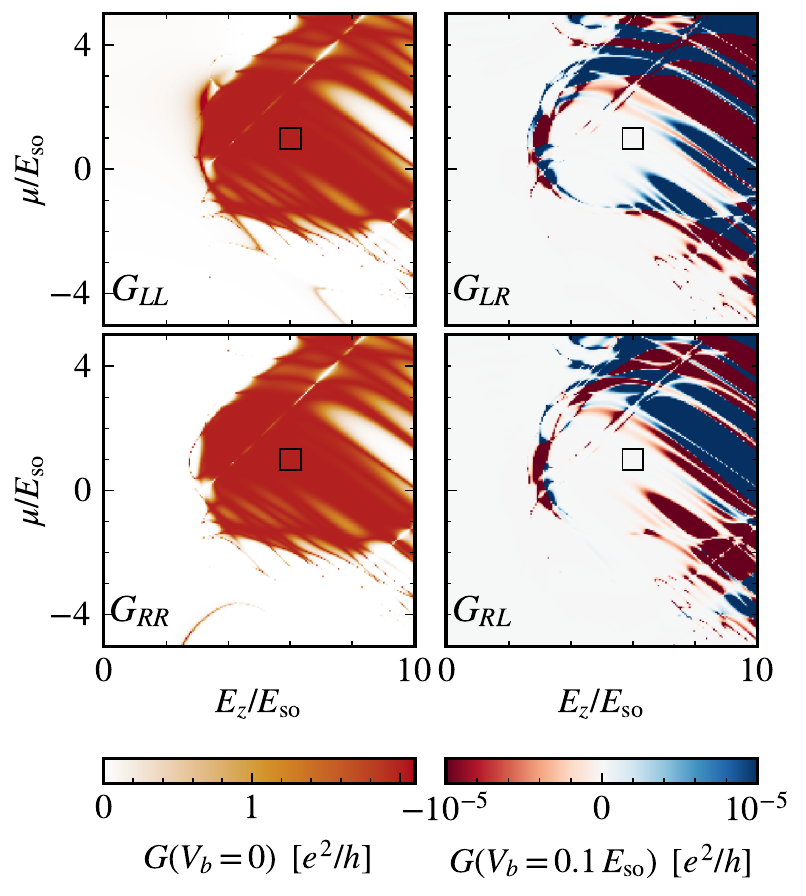}  
    \caption{\label{Fig:optgp}Elements of the conductance matrix between lead $L$ and $R$ as a function of the Zeeman field $E_z$ and chemical potential $\mu$ for the disordered wire with optimized gate voltages (see Fig.~\ref{Fig:opt}). The local conductances are shown at zero bias and the non-local conductances for a small bias voltage of $V_{\rm bias}=0.1\,\Eso$. Based on these conductance measurements, the topological gap protocol would indicate an extended topological phase with gap closing along the boundary. The black square indicates the point in the phase diagram at which the optimization has been performed.}
\end{figure}
It is crucial for a successful optimization to choose a metric that enhances the desired features when minimized. As disorder can induce trivial ABSs at zero energy, the metric cannot solely rely on the occurrence of zero-bias conductance peaks. We suggest the following metric 
\begin{align}
    \mathcal{M}(\bm{V}_g; E_z,\mu) &=
      -\tilde{G}_{LL}({V_{b}=0}) \tilde{G}_{RR}({V_{b}=0}) \,\frac{2\hat{\Delta}_{\rm gap}}{\Delta}  
      \notag\\
      &  
      \times \left|\ln \max_{E\in I_{E_z}} \sum_\alpha \tilde{G}_{\alpha\bar{\alpha}}(E_z=E)\right|^{-1} \ ,\label{Eq:metric}
\end{align}
which consists of the following contributions based on the ideas of the topological gap protocol \cite{Pikulin.2021}, as illustrated for the clean wire in Fig.~\ref{Fig:refmetric}:
\begin{enumerate}[(i)]
    \item%
            As a first contribution the local zero-bias conductance  $\tilde{G}_{\alpha\alpha}({V_{b}=0}) \equiv (h/e^2)\,G_{\alpha\alpha}(\bm{V}_g, V_b=0; E_z,\mu)$ at both loads is measured (left panels). Their product is large if there are localized zero-bias peaks at both ends, as it is the case for localized MZMs.  
    \item%
            As a second contribution, to ensure that the zero-bias peaks are not realized by simply closing the gap, we include an estimator for the transport gap $\hat{\Delta}_{\rm gap}(\bm{V}_g; E_z,\mu)$ obtained from a scan of the anti-symmetric part of the non-local conductance $G_{\alpha\bar{\alpha}}^{\rm asym} = [G_{\alpha\bar{\alpha}}(eV_b) - G_{\alpha\bar{\alpha}}(-eV_b)]/{2}$ over a range of bias voltages $V_b = V_R - V_L$ (black, vertical lines in right panels). This scan is performed by increasing the bias using a step size $\delta V_b=0.05\,\Eso$ until the signal peaks at $eV_b = \hat{\Delta}_{\rm gap}$, indicating extended state above the gap.  The rescale factor $2/\Delta$ balances the metric such that if the transport gap is the full proximity gap $\Delta$, the contribution to the metric is $2$ -- the same factor each zero-bias peak would contribute in (i) for ideal MZMs. 
    \item%
            The third contribution involves a scan of the non-local conductances $\tilde{G}_{\alpha\bar{\alpha}}(\bm{V}_g, V_b=0; E_z,\mu)$ over Zeeman fields in the interval $I_{E_z} = [E_z-5\,\Eso, E_z]$ (black, horizontal lines in right panels) with step size $\delta E_z=0.15\,\Eso$, where $\bar{\alpha}$ is the opposite lead of $\alpha$. The contribution to the metric is then given by the logarithm of the maximum of the sum of these non-local conductances, where taking the logarithm ensures that the metric is not dominated by this term as the non-local conductance can change by several orders of magnitude during optimization. This term is large if there is a gap-closing at a Zeeman field in the interval, which is the case if the transition to the topological phase occurs for a Zeeman field smaller than $E_z$.
\end{enumerate} 
Contribution (i) is in analogy to the first \emph{region of interest} (ROI) of the topological gap protocol while (iii) implements parts of the second phase of the protocol where the gapless part of the boundary of the ROI is determined \cite{Pikulin.2021}.

\section{Optimization of a one-dimensional Majorana wire}
We next revisit the Majorana wire of length $L=32.5\,\lso$ with strong onsite disorder $\sigma_{\rm dis}=25\,\Eso$ considered in the right panels of Fig.~\ref{Fig:ref}. For a Zeeman energy of $E_z=6\,\Eso$ and a chemical potential of $\mu=1\,\Eso$, we use the CMA-ES algorithm to optimize voltages of $N_g=50$ gates to minimize the metric Eq.~\eqref{Eq:metric} (for optimization with fewer gates see Appendix C). The CMA-ES algorithm is a derivative-free, population based machine learning algorithm \cite{Hansen.2016}: In each step of the optimization a population, i.e.~a set of $n_{\rm pop}$ candidate gate voltage configurations, is drawn from a multivariate normal distribution. Then the  metric is measured for each candidate in the population. The only information about the physical system that the algorithm needs is the value of the metric for each candidate, and based on the $n_{\rm pop}/2$ best candidates, the parameters of the algorithm that determine the search region are updated. The idea is that the search region from which the population is drawn can first expand to find candidates close to the minimum and then contract around the minimum. For a detailed description of the CMA-ES algorithm, we refer the reader to Ref.~\cite{Hansen.2016} and Appendix B, and details on applying the algorithm to gate array optimization can be found in Ref.~\cite{Thamm.2023}.

Fig.~\ref{Fig:opt} depicts the Majorana wave function, energy levels, and topological gap when using the optimized gate voltages found by the CMA-ES algorithm. The algorithm is capable of restoring localized MZMs such that the Majorana condition holds again (panel a). A scan of the topological gap through chemical potentials $\mu$ and Zeeman energies $E_z$ (panel b) reveals that optimization not only restores the topological phase at the point of optimization ($\mu=1\,\Eso$, $E_z=6\,\Eso$) but for an extended region (shown in red in panel b). In addition, the gap is restored, as can be seen from the comparison of the energy levels before optimization (panel c, red circles) and for the optimized voltages (panel c, blue crosses). These results indicate that the algorithm is indeed able to learn the disorder profile as it found gate voltages such that the potential profile created along the wire compensates the disorder.

We further show that the optimized wire passes the topological gap protocol as it lies in the ROI where zero-bias conductance peaks occur at both ends (Fig.~\ref{Fig:optgp}, left panels), and the non-local conductances (Fig.~\ref{Fig:optgp}, right panels) indicate that the boundary of the ROI is gapless, while there is a finite gap within the ROI.

\section{Inferring the topological invariant from the gap}
\begin{figure}[t!]
    \centering
    \includegraphics[width=8.6cm]{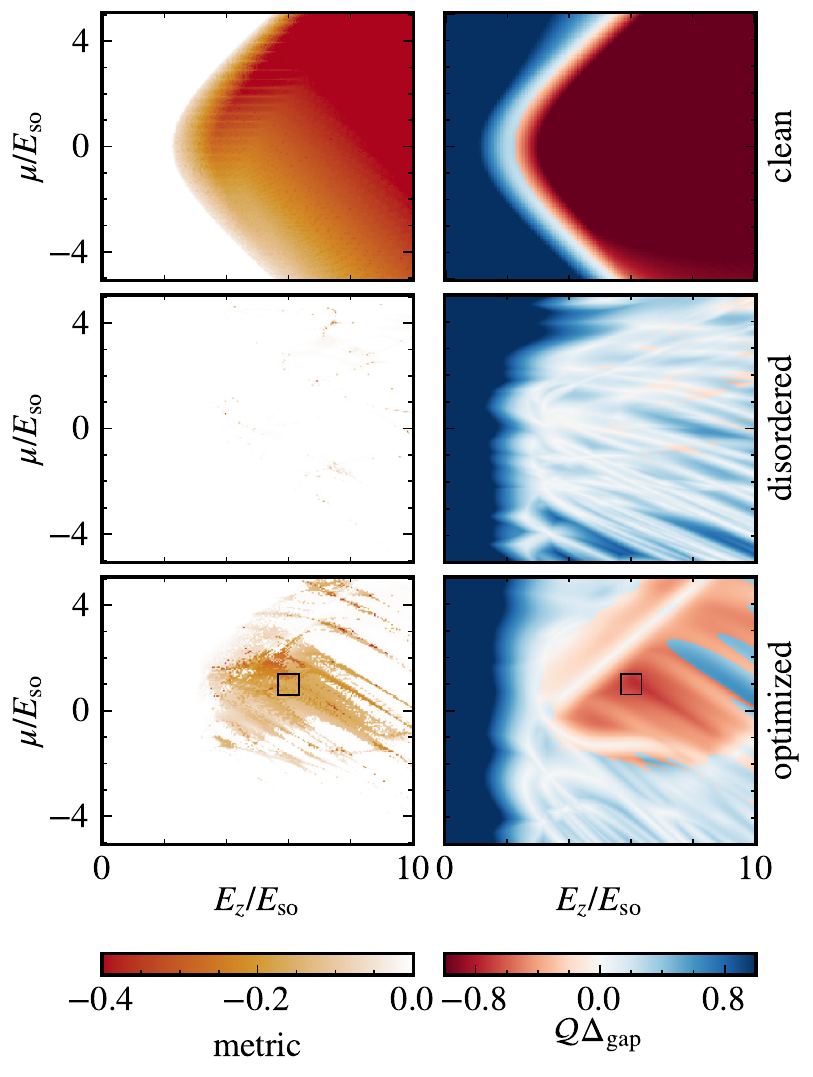}  
    \caption{\label{Fig:metricvstopgap}Comparison between the measurable metric (left panels) and the experimentally inaccessible topological gap $\mathcal{Q}\Delta_{\rm gap}$ (right panels) as a function of Zeeman energy $E_z$ and chemical potential $\mu$. The top panels depict a clean wire, the center panels a wire with strong disorder and zero voltage on all gates, and the lower panels show a wire with strong disorder and optimized gate voltages. The metric is small where the topological gap indicates a strong topological phase ($\mathcal{Q}=-1$ and large gap), and the topological phase can be accurately inferred from the values of the metric.}
\end{figure}
\begin{figure*}[t!]
    \centering
    \includegraphics[width=17.145cm]{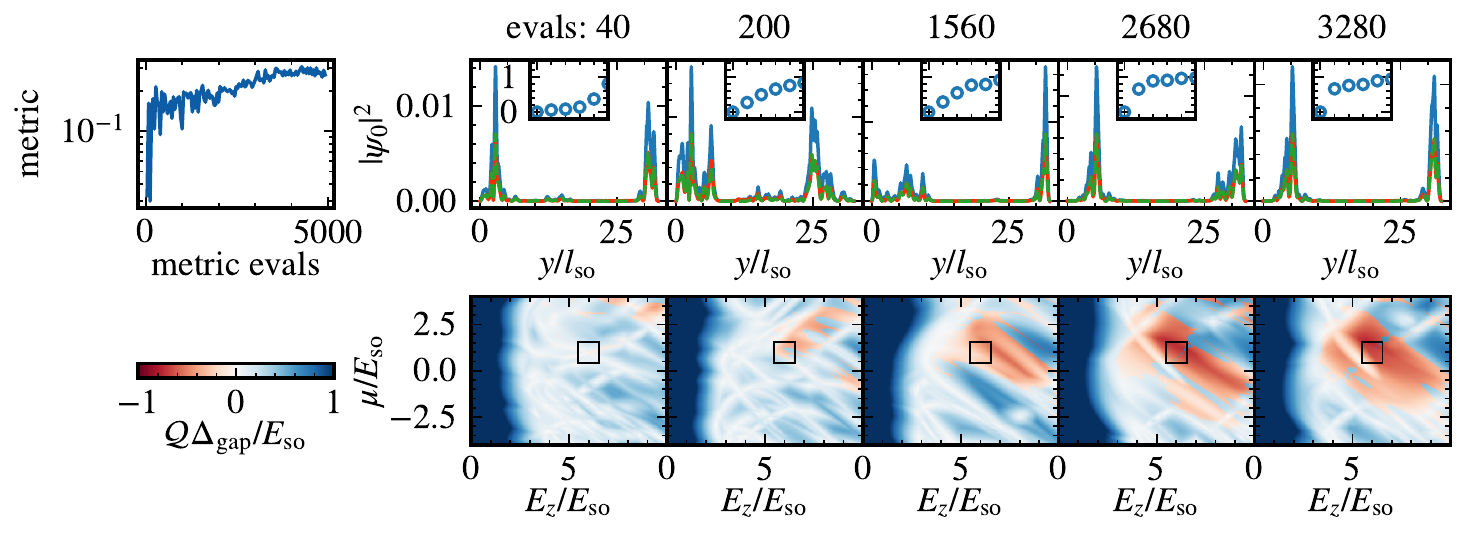}  
    \caption{\label{Fig:convergence}Convergence of the optimization algorithm for a wire of length $L=32.5\,\lso$ with strong onsite disorder with a strength of $\sigma_{\rm dis}=25\,\Eso$. The upper left-hand panel depicts the magnitude of the metric as a function of the number of metric evaluations. The other upper panels show the wave functions and the insets the lowest energy levels during several steps of the optimization before the algorithm has fully converged, and the lower panels show the corresponding topological gaps. To restore the localization of Majorana modes, only a few hundred metric evaluations are needed, and for obtaining an extended topological phase, a few thousand metric evaluations are sufficient even before full convergence of the CMA-ES algorithm.}
\end{figure*}
Since the desired properties of the system -- the topological phase and a large excitation gap -- are encoded in the topological gap $\mathcal{Q}\Delta_{\rm gap}$, one may assume that it is an ideal  metric for optimization. Unfortunately, however, the scattering invariant cannot be accessed experimentally, which disqualifies it for this purpose. To further motivate our metric, Eq.~\eqref{Eq:metric}, we show a comparison with the topological gap in Fig.~\ref{Fig:metricvstopgap}. For this, we calculate both quantities as a function of chemical potential $\mu$ and Zeeman energy $E_z$ for the clean wire (top panels), the disordered wire with $\sigma_{\rm dis}=25\,\Eso$, $\lambda_{\rm dis}=0$ (center panels), and the disordered wire with optimized gate voltages (bottom panels). We find astonishing agreement between our metric and the topological gap. It turns out that minimizing the metric leads to a topological phase with sizable excitation gap, and the absence of \emph{false positives} makes the optimization reliable for recovering the topological phase (see also Appendix D for more disorder realizations). 

\section{Convergence of the optimization}
Full convergence of the algorithm, such that for the best gate configurations at step $t$ and $t-1$ holds that $\|\bm{V}_g^{(t)}-\bm{V}_g^{(t-1)}\|< 10^{-8}\,\Eso$, may require tens of thousands of calculations (or measurements) of the metric, which would be experimentally very time-consuming. However, our aim is not to achieve full convergence, but to reliably compensate the disorder effects. We therefore consider Majorana wave functions, energy levels, and the topological gap at several steps during the optimization in Fig.~\ref{Fig:convergence}. The starting point for the optimization is again the disordered wire in the right panel of Fig.~\ref{Fig:ref} with zero voltage on all gates. It becomes apparent that the magnitude of the metric increases rapidly during the initial steps and that the slope subsequently flattens out (upper left panel). After only 40 metric measurements, localized MZMs are recovered at the point $(\mu,E_z)$. Already after 2680 metric measurements, an extended topological phase with considerable gap is obtained, and after 3280 measurements we no longer observe any significant improvements in the topological gap anymore, although the metric continues to increase slightly. From these observations, we conclude that complete recovery of the MZMs occurs significantly before formal convergence, so that a feasible number of much less than $5000$ evaluations of the metric is sufficient. In Appendix D, we show further optimizations for various disorder realizations after  3000 metric evaluations and find similar results.  In our simulations, one metric evaluation requires in total less than 100 conductance measurements for the scans along Zeeman field and bias voltage.

\section{Optimization in the presence of Andreev bound states}\label{Sec:ABSs}
The problem of a single conductance measurement is that it cannot distinguish between trivial ABSs pinned at zero energy and topological MZMs. This raises the question whether optimization of the metric, Eq.~\eqref{Eq:metric}, is able to ignore ABSs and tune the wire into the topological phase while simultaneously compensating for disorder. 

\begin{figure}[t!]
    \centering
    \includegraphics[width=8.6cm]{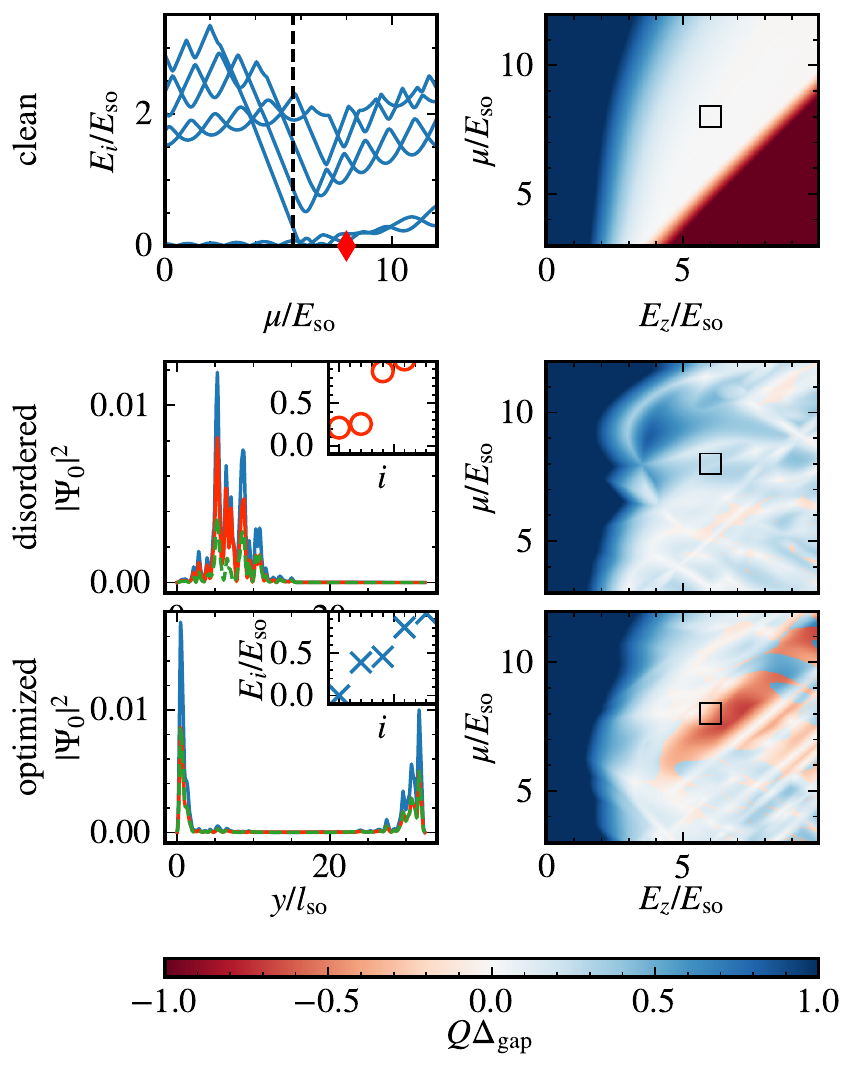}  
    \caption{\label{Fig:ABSb0}Results for a  wire with a smooth confinement potential that supports ABSs in the trivial phase. The upper panels show the lowest energy levels as a function of the chemical potential (left) and the topological gap (right) as a function of Zeeman field and chemical potential for a clean wire as a reference. Optimization is started in the trivial phase (red diamond), where a pair of ABSs is present in a clean wire when all gate voltages are set to zero. The center panels depict the wave function of the first level  $|\Psi_0|^2$ (blue, left panel) with their decomposition into hole (orange) and electron (green) components and the topological gap for the case of strong disorder $\sigma_{\rm dis}=50\,\Eso$, $\lambda_{\rm dis}=0.052\,\lso$, where all gate voltages are zero. The bottom panels show wave function and topological gap for using optimized gate voltages, where the mean voltage $b_0$ is included in the optimization. The insets in the wave function panels show the corresponding lowest energy levels. Optimizing the metric tunes the system to the topological phase -- ignoring the trivial ABSs. }
\end{figure}

By choosing a smooth confinement potential that slowly decays into the inside of the wire (see Appendix A for details), extended regions with a pair of ABSs at zero energy appear in the trivial phase \cite{Kells.2012} (white region in top panel of Fig.~\ref{Fig:ABSb0}), each localized at one end of the wire, thus giving rise to zero-bias conductance peaks at both ends \cite{Kells.2012}. 

As a starting point for optimization, we choose a chemical potential of $\mu=8\,\Eso$ and Zeeman energy $E_z=6\,\Eso$, which corresponds to the trivial phase with ABSs. In addition, we introduced strong disorder $\sigma_{\rm dis}=50\,\Eso$ with very short correlation length $\lambda_{\rm dis}=0.052\,\lso$, which causes the topological phase to vanish and destroys the localization of trivial ABSs (center panel in Fig.~\ref{Fig:ABSb0}). To allow the algorithm to tune out of the trivial phase into the topological phase, we here include the mean gate voltage $b_0$ in Eq.~\eqref{Eq:VjFourier} in the optimization. We find that the CMA-ES algorithm successfully compensates for the disorder while also ignoring the ABSs and tuning the wire into the topological phase (lower panel), resulting in localized MZMs.  
 
\section{Optimization of a two-dimensional Majorana wire}

\begin{figure*}[t!]
    \centering
    \includegraphics[width=17.145cm]{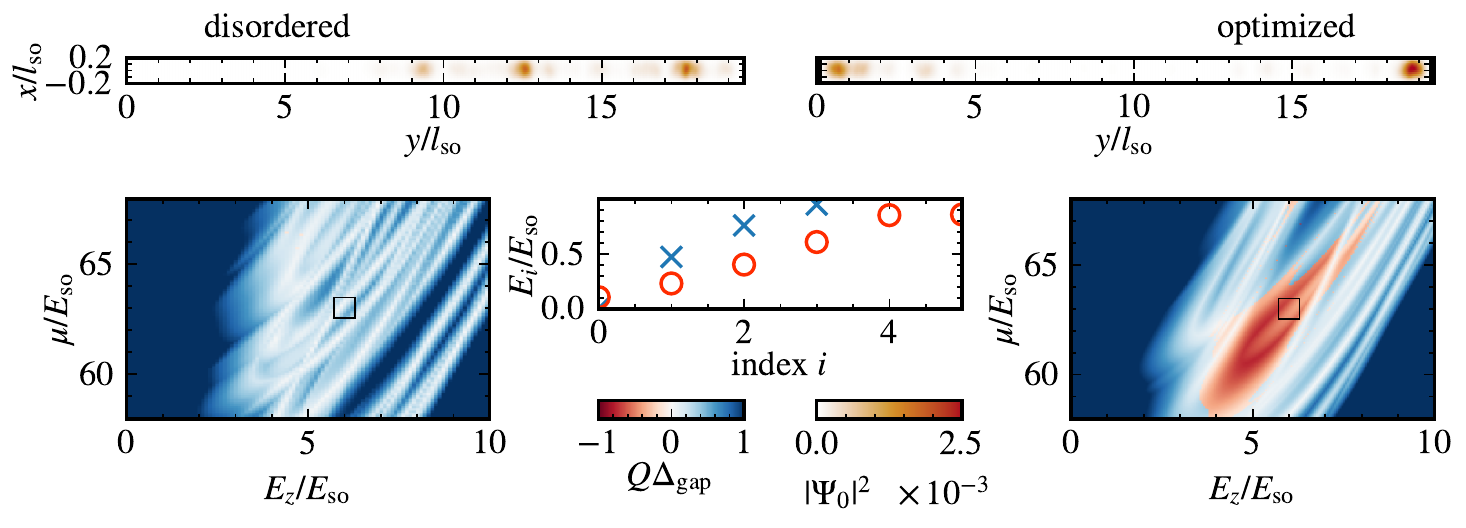}  
    \caption{\label{Fig:twodim}Optimization of a two-dimensional wire with strong disorder of strength $\sigma_{\rm dis}=120\,\Eso$ and short correlation length $\lambda_{\rm dis}=0.052\,\lso$ in the first topological phase for $\mu= 63\,\Eso$ and $E_z=6\,\Eso$. The upper panels show the wave functions $|\Psi_0|^2$ for the disordered wire with zero voltage on all gates (left) and for optimized gate voltages (right). The corresponding topological gaps $\mathcal{Q}\Delta_{\rm gap}$ are shown in the panels below. The lower center panel shows the lowest energy levels of the wire before optimization (red circles) and with optimized voltages (blue crosses). While the disorder completely destroys the localization of the wave function of the first level and the topological gap, both are restored when using optimized gate voltages.}
\end{figure*}
We next focus on a two-dimensional, rectangular wire of length $L_y=19.5\,\lso$ and width $L_x=0.39\,\lso$ described by the Hamiltonian
\begin{align} 
    \mathcal{H}^{2d}_{\rm wire} &= \tau_z \Bigg[ 
                -\frac{\hbar^2}{2m^*}(\partial_x^2+\partial_y^2)\sigma_0 - \mu\sigma_0 + \delta_{\rm dis}(x,y)\sigma_0
                \notag\\
                &-i\hbar\alpha_R(\sigma_x\partial_y-\sigma_y\partial_x) 
                + V_g(x,y)\sigma_0
              \notag\\
              & +V_{\rm conf}(y) \sigma_0
                \Bigg]  +\frac{\mu_B g B_z}{2}\tau_0\sigma_z + \Delta\tau_x\sigma_0
                \ . \label{Eq:Hwire2d} 
\end{align}
Here, the Landé factor is given by $g=-14.9$ \cite{Winkler.2019}, and if not stated otherwise, the same parameters as for the one-dimensional wire are chosen. When discretizing the Hamiltonian on a two-dimensional lattice with spacings $a_x=a_y=0.026\,\lso$, we include the orbital effect of the magnetic field by adding Peierls phases $\e^{-i e/\hbar \int_{\bm{r}_1}^{\bm{r}_2} \bm{A}\cdot d\bm{r}}$ to the hoppings from site $\bm{r}_1$ to $\bm{r}_2$. We choose the vector potential $\bm{A}$ such that it is given by $-B_z x \bm{e}_y$ away from the ends and smoothly vanishes over a distance $L_x/2$ towards the short ends  of the wire in order to conserve the supercurrent in the wire \cite{Thamm.2023}.

The clean two-dimensional wire is in the first topological phase for a chemical potential of $\mu=63\,\Eso$ and a Zeeman energy $E_z=\mu_B g B_z/2= 6\,\Eso$, which we choose as parameters for the optimization. Adding strong disorder with strength $\sigma_{\rm dis}=120\,\Eso$ and a short correlation length $\lambda_{\rm dis}=0.052\,\lso$ completely destroys MZMs (Fig.~\ref{Fig:twodim}, upper left panel) and the topological phase (lower left panel) similarly to the one-dimensional case.  Again optimization of $N_g=50$ gates allows restoring localized MZMs (upper right panel) and an extended topological phase (lower right panel).

\section{Conclusions}
In this study, we explored the machine learning optimization of a gate array placed in close proximity to a strongly disordered Majorana wire. To optimize the system using the CMA-ES algorithm, we introduced a metric based on the topological gap protocol. This metric allowed for the optimization of a grounded wire connected to two leads, eliminating the need for interferometry and Coulomb blockade. By minimizing the metric, the CMA-ES algorithm effectively restored the localized Majorana zero modes (MZMs), extended the topological phase, and reopened the excitation gap, even in cases where they were completely destroyed by the disorder. Remarkably, the algorithm demonstrated the capability to disregard trivial Andreev bound states (ABSs) and to tune the wire into the topological phase while simultaneously mitigating the effects of disorder. Furthermore, we demonstrated that the required number of measurements for the optimization process is experimentally feasible, and even with a modest number of gates (around 20-50), substantial improvements can be achieved. Notably, gate arrays of this scale can already be constructed using standard electron beam lithography and aluminum gates isolated by native oxide \cite{Mills.2019,PoeschlPhD.2022}.

\begin{acknowledgments}  
    This work has been funded by the Deutsche Forschungsgemeinschaft (DFG) 
    under Grants No.~RO 2247/11-1 and No.~406116891 within the Research 
   Training Group RTG 2522/1.\\ 
\end{acknowledgments}

\renewcommand{\theequation}{A\arabic{equation}}
\setcounter{equation}{0} 
\section*{APPENDIX A: Details on disorder and confinement potentials} 
    To obtain correlated disorder potentials, we first draw random numbers $\delta$ with 
	standard deviation $\sigma_{\rm dis}$ from a normal distribution. A finite  correlation length $\lambda_{\rm dis}$ can be included by damping high Fourier modes according to \cite{Thamm.2023}
	\begin{align}
		\delta_{\rm dis}(y) = \mathcal{F}^{-1}\left[
				\e^{-|q|\lambda_{\rm dis}} \mathcal{F}[\delta(y)]
				\right]\ .
	\end{align}
    For most of the one-dimensional wires considered here, we choose onsite disorder $\lambda_{\rm dis}=0$, which is the most challenging type of disorder to compensate using gate potentials \cite{Thamm.2023}. 
	For the two-dimensional wire, we consider the case $\lambda_{\rm dis}=0.052\,\lso$, which is a very short  correlation length.

    A confinement potential $V_{\rm conf}$ separates the wire from the leads where we lower the potential by $V_{\rm lead}=100\,\Eso$ to ensure that both spin species are present at the Fermi level in the leads. Except for Sec.~\ref{Sec:ABSs}, we consider a steep confinement of the form $V_{\rm conf}(y) = V_{\sigma,V_0}(y-y_0) + V_{\sigma,V_0}(y-L+y_0)$ with $V_{\sigma,V_0}(y)=V_0\exp{-y^2/(2\sigma^2)}$, $\sigma=0.1\,\Eso$, and $V_0=65\,\Eso$. Here, $y_0=\sqrt{2\sigma^2\ln 2}$ such that the potential maximum is moved a distance $y_0$ into the wire and the potential has decayed to $V_0/2$ at the wire lead interface.

    In Sec.~\ref{Sec:ABSs}, we use a smooth confinement potential in order to produce a region of ABSs at zero energy in the trivial phase \cite{Kells.2012,Cayao.2021}. The potential is made up of a narrow Gaussian peak with decay length $\sigma_1=0.1\,\lso$ and a wide peak with $\sigma_2=6\,\lso$ continuously matched at $(y_1+y_0,E_s)$. At the left lead, the potential is given by 
    \begin{align}
        V(y) &=\begin{cases}  
            V_{\sigma_1,V_0+V_{\rm lead}}(y-y_0)-V_{\rm lead} 
                &  y\leq 0\\
            V_{\sigma_1,V_0}(y-y_0)                               
                &  0 < y < y_1\\	
            V_{\sigma_2,V_0}(y-y_0-y_1+y_2)                        
                &  y \geq y_1\\
        \end{cases}  \notag\\
        y_0 &= \sqrt{2\sigma^2\ln 2}\\
        y_j  &=\sqrt{2\sigma_j^2\ln(V_0/E_s)} \ . \label{Eq:ABSPot}
    \end{align}
    We choose $E_s=10\,\Eso$, $V_0=65\,\Eso$, and $V_{\rm lead}=100\,\Eso$.


\begin{figure}[t!]
    \centering
    \includegraphics[width=8.6cm]{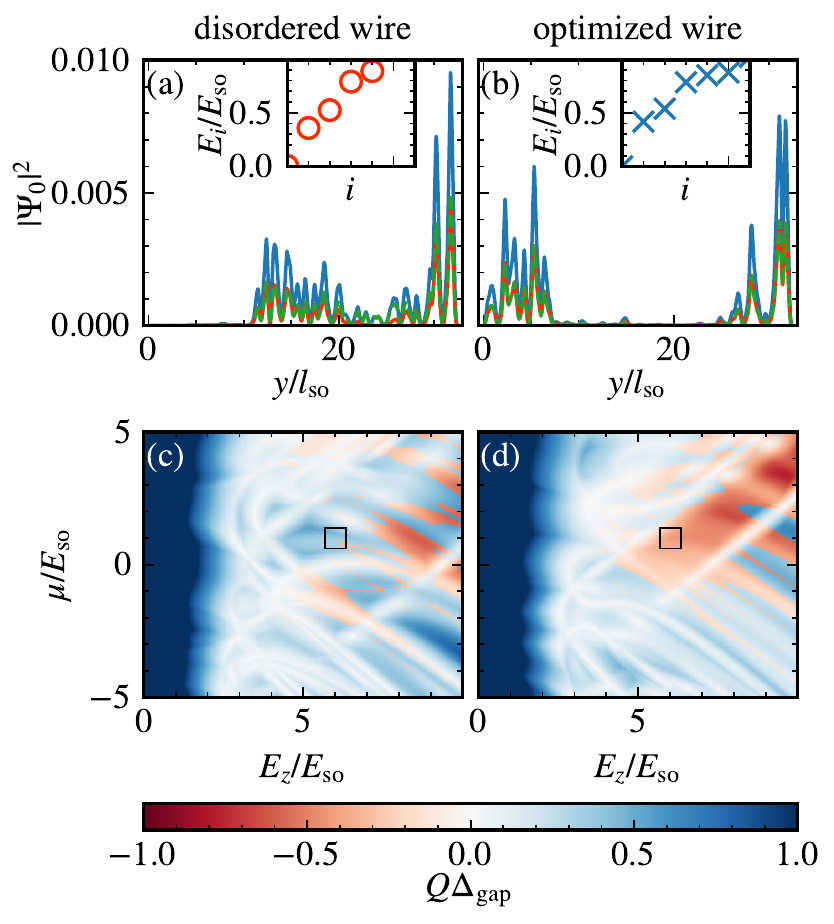}  
    \caption{\label{Fig:20gates}Optimization of only 20 gates in proximity to a  Majorana wire of length $L=32.5\,\lso$ with chemical potential $\mu=1\,\Eso$ and Zeeman energy $E_z=6\,\Eso$. Panels (a) and (b) depict the wave functions $|\Psi_0|^2$ (blue) of the first level together with the electron $|u_0|^2$ (green) and hole wave function $|v_0|^2$ (orange) for the disordered wire with short correlations ($\sigma_{\rm dis}=25\,\Eso$, $\lambda_{\rm dis}=0.05\,\lso$) and for the same wire with optimized gates, respectively. Panels (c) and (d) show the corresponding topological gaps $\mathcal{Q}\Delta_{\rm gap}$. Optimization of only 20 gates restores an extended topological phase and the localization of the MZMs.}
\end{figure}
%
\begin{figure*}[t!]
    \centering
    \includegraphics[width=17.0cm]{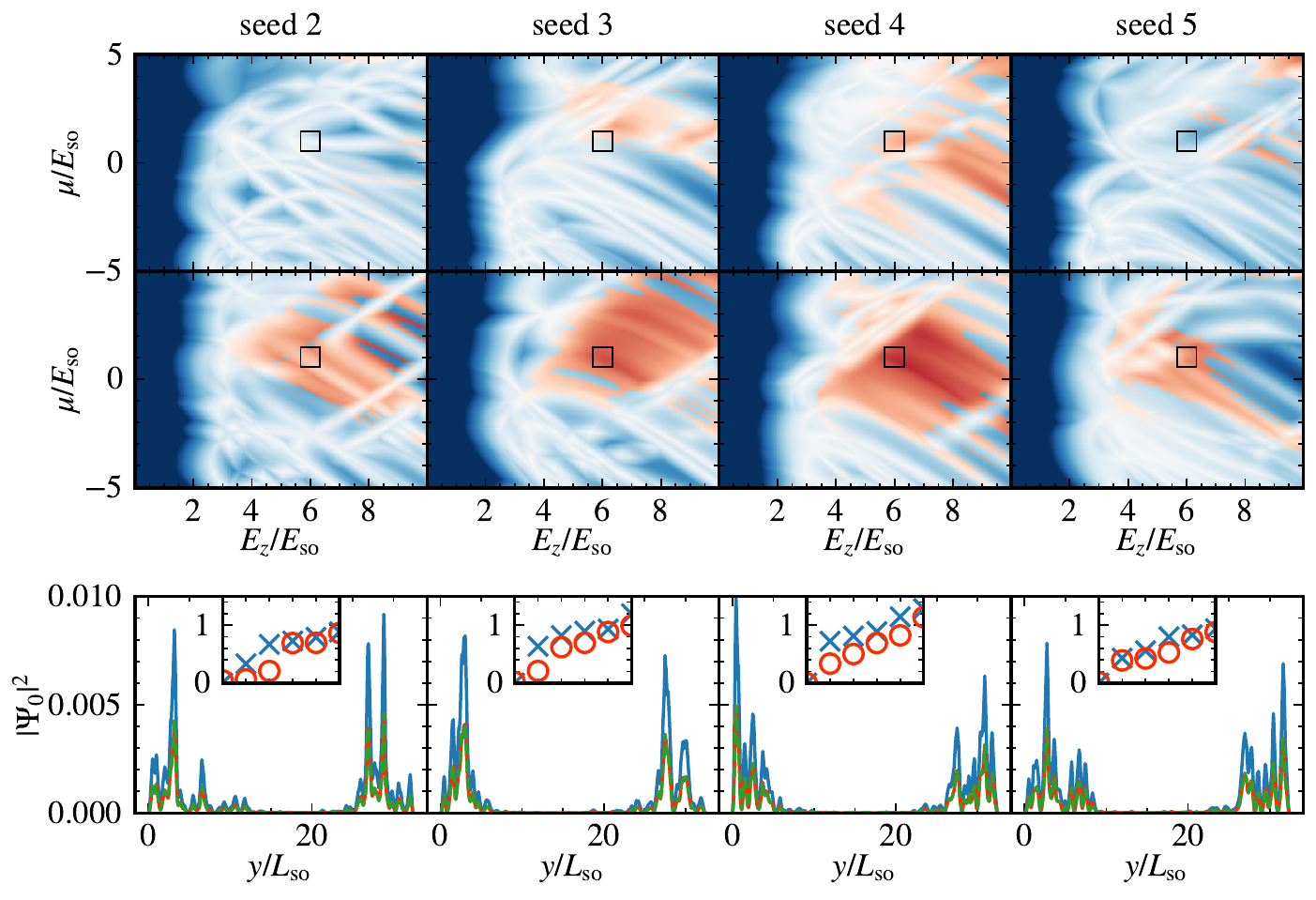}  
    \caption{\label{Fig:seedDep}Optimization results for a one-dimensional wire of length $L=32.5\,\lso$ with different realizations of the onsite disorder with strength $\sigma_{\rm dis}=25\,\Eso$ (by using different seeds of the random number generator for drawing disorder profiles; cf.~Fig.~\ref{Fig:opt} for seed 1). Top panels depict the topological gap $\mathcal{Q}\Delta_{\rm gap}$ before optimization with zero voltage on all gates. The center panels show the topological gap for optimized gate voltages after 3000 metric evaluations (75 CMA-ES steps with population size 40). The lower panels show the corresponding wave functions after optimization $|\Psi_0|^2$ (blue), the electron component $|\bm{u}_0|^2$ (green), and the hole component $|\bm{v}_0|^2$ (orange). The insets depict the lowest energy levels before optimization (red circles) and after optimization (blue crosses). Optimization yields an extended topological phase for all considered disorder realizations.}
\end{figure*}

\renewcommand{\theequation}{B\arabic{equation}}
\setcounter{equation}{0} 
\section*{APPENDIX B: Details on CMA-ES optimization} 
    In each step $t$ of the CMA-ES algorithm \cite{Hansen.2016}, we draw a population of $n_{\rm pop}=40$ gate voltage configurations -- or rather their Fourier components $(\bm{a}, \bm{b})$ -- from a multivariate normal distribution $\mathcal{N}\big((\bm{a}^{(t)},\bm{b}^{(t)}), (\sigma^{(t)})^2{\bf \sf C}^{(t)}\big)$. Here, $(\bm{a}^{(t)},\bm{b}^{(t)})$ is the mean of step $t$ obtained as a weighted average of the $n_{\rm pop}/2$ best candidates of step $t-1$, i.e.~the candidates with the smallest value for the metric. The other parameters are the step size $\sigma^{(t)}$ and the correlation matrix ${\bf \sf C}^{(t)}$ adjusted by the CMA-ES algorithm to find candidates close to the minimum  and contract the search region around it \cite{Hansen.2016}. As initial parameters, we set $\sigma^{(0)}=1.0$ , ${\bf \sf C}^{(0)} = \mathds{1}$, and $(\bm{a}^{(0)},\bm{b}^{(0)}) = 0$, i.e.~zero voltage on all gates. We use the mature \texttt{python} implementation \texttt{pycma} \cite{Hansen.2021} with formal convergence criteria \emph{topfun}$\;=10^{-15}$, \emph{tolfunhist}$\;=10^{-8}$, and \emph{tolx}$\;=10^{-5}\,E_{\rm so}$ and otherwise default parameters to perform the CMA-ES optimization computations.

\renewcommand{\theequation}{C\arabic{equation}}
\setcounter{equation}{0} 
\section*{APPENDIX C: Optimization with a reduced number of gates} 
In Ref.~\cite{Thamm.2023} a sweet spot in the number of gates for optimization of Majorana wires was identified as about $1.5$-$2$ gates per $1\,\lso$ of wire length. For the wires of length $L=32.5\,\lso$ considered here, using 50 gates is within this range. However, the larger the number of gates, the more challenging is the implementation for larger scale devices. We therefore also consider optimization using only 20 gates. In Fig.~\ref{Fig:20gates}, we show optimization for strong disorder $\sigma_{\rm dis}=25\,\Eso$ with short correlation length $\lambda_{\rm dis}=0.05\,\lso$ -- much smaller than the extension of an individual gate ($1.625\,\lso$). While disorder destroys the localization of the MZMs and most of the topological phase (panels a and c), optimization of 20 gates is capable of restoring an extended topological phase and the MZM localization (panels b and d). These results indicate that a much smaller number of gates than the sweet spot is already sufficient to improve Majorana devices. In case of larger correlation lengths or smaller disorder strength, we expect that optimizing an even smaller number of gates  yields significant improvements for the performance of Majorana devices.

\renewcommand{\theequation}{D\arabic{equation}}
\setcounter{equation}{0} 
\section*{APPENDIX D: Dependence on the disorder realization} 
To demonstrate that the optimization reliably restores the topological phase, we consider several disorder realizations for onsite disorder with strength $\sigma_{\rm dis}=25\,\Eso$ by using different seeds for the random number generator (seed 1 is used in the main text). In the top panel of Fig.~\ref{Fig:seedDep}, we show the topological gap for the disordered wires where all gate voltages are set to zero. At the point of optimization (black square), disorder destroyed the topological phase in four out of five cases (including the main text results), which is a realistic amount of disorder \cite{Aghaee.2022}. 

For a realistic test, we stop optimization after 3000 metric evaluations and use the best gate voltages at this point. In all cases, we find an extended topological phase (center panel) and localized MZMs (lower panel) when using the optimized gate voltages.

%

\end{document}